%% file: main.tex
\documentclass[sigplan,twocolumn,nonacm]{acmart}
\acmSubmissionID{167}
\renewcommand\footnotetextcopyrightpermission[1]{}
\AtBeginDocument{%
  }
    
\usepackage{xcolor}
\usepackage{comment}
\usepackage[commandnameprefix=ifneeded]{changes}
\usepackage{algorithm}
\usepackage{algpseudocode}
\usepackage{multirow}
\usepackage{tablefootnote}
\usepackage{threeparttable}
\usepackage{makecell}
\usepackage{array}
\usepackage{subcaption}
\usepackage{hyperref}
\usepackage{pifont}
\usepackage{caption}
\usepackage{float}
\usepackage{stfloats}

\newcommand{\ignore}[1]{}

\usepackage{physics}
\usepackage{tikz}
\usepackage{tcolorbox}
\definecolor{OliveGreen}{HTML}{135D66}
\newtcolorbox{hintbox}[2][]
{
  colback  = OliveGreen!5,
  boxsep=2pt,
  boxrule=0.2mm,
  titlerule=0.2mm,
  width=\dimexpr\columnwidth\relax, 
  coltitle = blue!20!black,
  title    = #2,
  #1,
}
\definecolor{lblue}{HTML}{F6FBFC}

\acmSubmissionID{167}

\usepackage{enumitem}
\setcopyright{none} 
\copyrightyear{2026}
\acmYear{2026}
\acmDOI{XXXXXXX.XXXXXXX}

\acmISBN{978-X-XXXX-XXXX-X/XX/XX}

\input{sections/macros}
\begin{document}

\title{\papername: GPU-Variability-Aware Expert-to-GPU Mapping\\For Mixture-of-Experts Models}

\author{Sourish Wawdhane}
\email{sourishw@utexas.edu}
\affiliation{%
  \institution{The University of Texas at Austin}
  \city{}
  \state{}
  \country{}
}

\author{Avinash Kumar}
\email{avinkumar@utexas.edu}
\affiliation{%
  \institution{The University of Texas at Austin}
  \city{}
  \state{}
  \country{}
}

\author{Poulami Das}
\email{poulami.das@utexas.edu}
\affiliation{%
  \institution{The University of Texas at Austin}
  \city{}
  \state{}
  \country{}
}




\input{sections/0_Abstract}

\maketitle

\input{sections/1_Introduction}
\input{sections/2_Background}

\input{sections/3_Design}

\input{sections/4_Methodology}
\input{sections/5_Results}
\input{sections/6_Discussion}

\input{sections/7_Related}

\input{sections/8_Conclusion}
\input{sections/9_Acknowledgements}


\bibliographystyle{unsrt}
\bibliography{references}
 \clearpage
 \appendix
 \input{sections/A0_VariabilityData}
 \input{sections/A1_Algorithm}
 \input{sections/A2_AdditionalResults}

\end{document}

%% file: sections/macros.tex
\usepackage{xspace} 

\newcommand{\papername}{\textsc{GEM}\xspace}

\newcommand{\moe}{MoE\xspace}

\renewcommand{\trace}{trace\xspace}
\renewcommand{\Trace}{Trace\xspace}

\usepackage{xcolor}
\usepackage{xspace}
\usepackage{tikz}

\newcommand*{\circled}[1]{\tikz[baseline=-0.7ex]{
    \node[shape=circle, fill=black, inner sep=1pt, minimum size=1.0em] {\textbf{\color{white}\scriptsize #1}};}}

%% file: sections/0_Abstract.tex
\begin{abstract}

Mixture-of-Expert (MoE) models enable efficient inference by employing smaller \textit{experts} and activating only a subset of them per token. 
MoE serving engines distribute experts across multiple GPUs and route tokens to appropriate GPUs at inference time based on experts activated. They process tokens in lock-step fashion, where tokens within a batch must finish processing before proceeding to the next layer. This synchronization barrier acts as a critical bottleneck because the performance of MoE models is limited by the \textit{straggler} GPU that finishes last. Stragglers emerge when too many heavily used experts are placed on the same GPU or the slowest GPU. While prior works place experts that balance token loads across GPUs, they \textit{all} overlook GPU variability and often place highly used experts on the slowest GPUs. 

We propose \textit{\papername, \underline{G}PU-variability-aware \underline{E}xpert \underline{M}apping}, a framework for GPU variability-aware expert to GPU mapping for MoE models. \papername exploits two insights. \textit{First}, we must place experts such that each GPU receives
\textit{non-uniform} token loads based on their variability and they
all finish processing a layer at about the same time. Our studies show that there are two types of experts: \textit{consistent} that are used most of the time and \textit{temporal} that are often used together for the remaining time. Our \textit{second} insight is that we must place simultaneously used consistent and temporal experts on different GPUs and avoid placing them on slower GPUs to reduce slowdown. \papername gathers the variability profile of GPUs for each model and task and uses the token load distributions per task to map experts to GPUs. Our experiments show that \papername improves end-to-end latency by 7.9\% on average and by up to 16.5\% compared to the baseline. 
\end{abstract}

%% file: sections/1_Introduction.tex
\section{Introduction}
\label{sec:introduction}

Feed-forward-networks (FFNs) account for up to two-thirds of per token compute costs in Large Language Models (LLMs)~\cite{pei2025cmoe,geva2021transformerfeedforwardlayerskeyvalue}. Mixture of Experts (\moe) models are variants of LLMs that lower these costs by replacing each FFN with smaller sub-networks, called \textit{experts}, and use only a subset of them to process each token. For example, Mixtral 8x7B~\cite{mixtral} activates only 2 out of its 8 experts for each token and yet, achieves accuracy comparable to the dense Llama-2 70B model while using $\sim$\!5$\times$ fewer floating point operations per token~\cite{mixtral}. This makes \moe models very attractive for high-throughput and low-latency inference, leading to their increased adoption in real-world serving environments in recent years~\cite{deepseekai2025deepseekv3technicalreport, qwen3, llama4, hunyuan, deepseekr1, muennighoff2024olmoe, dai2024deepseekmoe}.

\begin{figure*}[t]
    \centering
    \includegraphics[width=0.98\textwidth]{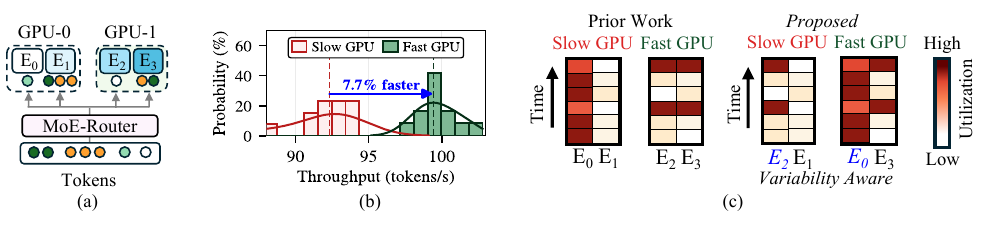}
    \caption{(a) \moe models are deployed such that experts are distributed across multiple GPUs and a router selects expert(s) to process each token. (b) GPUs exhibit performance variability. (c) \papername places experts on GPUs such that a fast GPU processes proportionally more tokens in the same time as a slow GPU and avoids placing heavily used experts on slower GPUs.}
    \label{fig:fig1}
\end{figure*}

The performance of MoE models is severely limited by \textit{stragglers}. Figure ~\ref{fig:fig1}(a) shows how \moe models use expert parallelism, where the experts are distributed across multiple GPUs; and a router determines the experts to be activated for each token and sends the token to the GPUs hosting those experts. 
However, due to lockstep execution, a batch can proceed to the next layer only after the tokens corresponding to all its requests have been processed in the current layer. This synchronization barrier creates a major bottleneck because a batch must wait for the slowest GPU or the \textit{straggler} to finish execution before proceeding. The problem worsens because experts are not used uniformly; some are used more often than others and the most frequently used experts vary across layers~\cite{fedus2022switch, mixtral}. For instance, our experiments using the Qwen3-235B model on the ShareGPT dataset show that the most frequently used expert is used 4.2$\times$ more often than the least used expert. Reducing the impact of straggler GPUs is critical to improve the performance of \moe models. 

A GPU becomes a straggler if it hosts too many highly utilized experts and/or if it is inherently slower than the others due to hardware variability. While prior works balance the mapping of highly used experts by introducing artificial uniformity in expert utilization which degrades accuracy~\cite{fedus2022switch, lepikhin2020gshard} or by dynamically re-distributing experts across GPUs incurring substantial memory overheads~\cite{deepseekai2025deepseekv3technicalreport, craft2025}, 
they all overlook the inherent \textit{variability} in the GPU hardware. 
In reality, modern GPUs exhibit substantial performance variability due to process variation, environmental factors, and runtime power management~\cite{sinha2022not, jain2024palvariabilityawarepolicyscheduling}. Our experiments across 128 NVIDIA L40 GPUs show that inference throughput differs by up to 27.7\% between the fastest and the slowest GPUs and Figure \ref{fig:fig1}(b) shows that even within a single 8-GPU node, the spread persists at 7.7\% over a week. Even if we were to perfectly balance the load across the GPUs, the problem of stragglers would still persist due to fundamental limitations in the hardware. The severity of the problem increases in real-world deployment scenarios using dozens of GPU~\cite{deepseekai2025deepseekv3technicalreport, llama4, hunyuan} because the likelihood of encountering non-uniformity in the hardware and the probability of using slower GPUs increase. Straggler GPUs become the steady-state bottleneck because even a single slow GPU stalls every layer of every forward pass throughout the deployment. \textit{Ideally}, \moe model serving frameworks must \textit{map} experts on GPUs such that the impact of stragglers stemming from both non-uniform expert utilization and GPU variability is minimized.


In this paper, we propose \textit{\papername, \underline{G}PU-variability-aware \underline{E}xpert \underline{M}apping}, a framework for high-throughput low-latency \moe model serving via GPU variability-aware expert mapping. A natural strategy to enable GPU-variability-aware expert mapping would be to simply place the most heavily used experts on the fastest GPUs. But this approach does not work for two reasons. \textit{First}, the fastest GPU in a node is often only modestly faster than the slowest one. This limited gap constrains the extra token load that the fastest GPU can process. Naively placing too many frequently-used experts on the fastest GPU may exceed this limit, causing it to become a straggler itself.
\textit{Second}, the most heavily-used experts fall into two categories: \textit{consistent} experts that are active most of the time, and \textit{temporal experts} that are highly utilized in the remaining time (small fraction of the total time). Unfortunately, while metrics like \textit{average token load} over a span of time, used in prior works~\cite{dai2024deepseekmoe, fedus2022switch, wang2024auxiliary}, suffice for detecting consistent experts, they cannot detect temporal experts.

\papername addresses these challenges by exploiting two key insights. \textit{First}, rather than trying to equalize token loads across GPUs, \papername places experts such that GPUs receive non-uniform token loads based on their performance variability and the faster GPUs only process proportionally more tokens than the slower ones, allowing all GPUs to complete processing for each layer at about the same time. For example, on an NVIDIA 8xL40, the fastest GPU processes 14\% more tokens than the slowest at the same latency. 
This enables \papername to transform hardware variability from a liability to a mapping constraint that the framework exploits to improve performance.
\textit{Second}, \papername optimizes for both consistent and temporal experts. By observing token loads at a much finer granularity rather than relying on average utilization, \papername detects both types of experts. 
To optimally place the temporal experts, \papername exploits our observation that many of these experts are often used together. In other words, their utilization is \textit{correlated}. Figure \ref{fig:fig1}(c) illustrates how \papername ensures simultaneously used consistent and temporal experts are placed on different GPUs and that they are not placed on slower GPUs (which is GPU-0 in the illustration).

\papername employs a four-step process. First, \papername captures the number of tokens that each expert receives at each step during online inference to balance the mapping of both temporal and consistent experts. Second, \papername profiles performance variability across GPUs by measuring per-GPU \moe-layer latency as a function of token load. However, building these profiles at a fine granularity is non-trivial because it can take hours during which the GPUs are unavailable for inference, whereas coarse sampling under-samples the staircase pattern of \moe latency, where latency jumps sharply at certain token counts. To address this, \papername samples token counts only at tile boundaries, reducing profile time from hours to minutes. Third, \papername must identify an expert-to-GPU mapping that minimizes straggler latencies caused by both expert imbalance and performance variability. However, the expert mapping space grows combinatorial in the number of experts and GPUs. So, \papername runs a heuristic-based iterative search to refine mappings. Finally, GEM loads each expert's weights onto its assigned GPU at model load time and continues to use it throughout the deployment.


Our evaluations across five \moe models and representative tasks show that \papername improves end-to-end latency by 7.9\% on average and by up to 16.5\% compared to the baseline.

\noindent Overall, this paper makes the following contributions:
\begin{enumerate}[leftmargin=0cm,itemindent=.5cm,labelwidth=\itemindent,labelsep=0cm,align=left, itemsep=0.07 cm, listparindent=0cm, topsep=0.3 cm]
    \item We observe that performance of MoE models is limited by stragglers resulting from too many heavily used experts placed on the same GPU and inherent hardware variability. 

    \item We propose {\em \papername, \underline{G}PU-variability-aware \underline{E}xpert \underline{M}apping}, a framework for high-throughput low-latency \moe model serving via GPU-variability-aware expert to GPU mapping.
    
    \item We place experts such that the faster GPUs only process proportionally more tokens than the slower ones, allowing all GPUs to complete processing at about the same time. 

    \item We identify two types of experts: \textit{consistent} (used more often and easy to detect) and \textit{temporal} (used infrequently and hard to detect). We design \papername such that both of them can be detected and the experts that are simultaneously used are placed on different GPUs and avoid the slower GPUs. 
\end{enumerate}

%% file: sections/2_Background.tex
\ignore{
\begin{figure*}[htp]
    \centering
    \includegraphics[width=1\linewidth]{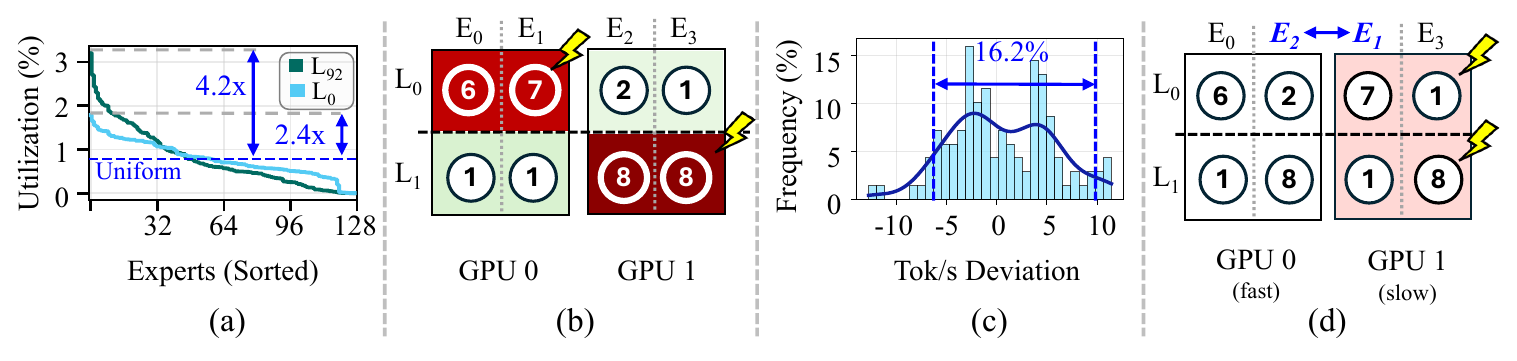 }
    \caption{(a) (Make bigger) Expert utilization ($\%$) for Layer-0 and Layer-92 of Qwen3-235B on ShareGPT~\cite{sharegpt}. Both layers have a subset of experts with high utilization, while the majority of experts remain underutilized. (b) Expert imbalance leads to token imbalance at the GPU level, creating straggler GPUs that stall inference at each \moe layer's synchronization point. (c) Modern GPUs exhibit significant performance variability. We profile throughput variation across NVIDIA 8xL40 nodes. (d) Even under balanced token processing, performance variability across GPUs creates straggler GPUs that take longer to process experts. }
    \label{fig:background_figure}
\end{figure*}
}
\newpage
\section{Background and Motivation}
\label{sec:background}




\subsection{Mixture-of-Expert Models}
\label{moe-background}
Mixture-of-Expert (MoE) models are an emerging class of LLMs that replace dense feed forward networks with multiple smaller networks, called ``experts"~\cite{jacobs1991adaptive, shazeer2017outrageously}. 
Each expert specializes in different linguistic representations~\cite{dai2024deepseekmoe, muennighoff2024olmoe} that differ across layers. For example, one expert may specialize in coding-related tasks, while another specializes in punctuation. At inference time, a router selects a subset of experts for each token~\cite{fedus2022switch, shazeer2017outrageously}. For instance, the Qwen3-235B MoE model uses only 8 out of 128 experts available per token while attaining high accuracy across most tasks ~\cite{qwen3}. Thus, \moe models drastically reduce computational costs of token generation, improving throughput and latency. Consequently, they are becoming increasingly prevalent~\cite{cai2025survey, mixtral, deepseekai2025deepseekv3technicalreport, deepseekr1, qwen3, fedus2022switch, lepikhin2020gshard}.

Modern inference engines, like vLLM~\cite{kwon2023efficient}, distribute experts across GPUs, a deployment strategy called \textit{expert parallelism}. Each \moe layer's router selects experts for each token and dispatches it to the GPU(s) hosting those experts. 
As tokens in a batch are processed in a lockstep manner, all experts \textit{must} finish processing before the batch can proceed to the next layer. This creates a synchronization barrier at each layer, where the last GPU to finish computation, called the \textit{straggler}, determines the effective latency. Minimizing stragglers is critical for performance~\cite{hwang2025capacity, harmoeny2025, go2025moetuner, yang2026libra, craft2025, li2023lina, li2025speculativemoe, luo2025occult, huang2024toward, deepseekai2025deepseekv3technicalreport}.

\subsection{Straggler Problem With Expert Parallelism}

\begin{figure}[htp]
    \centering
    \includegraphics[width=1.0\linewidth]{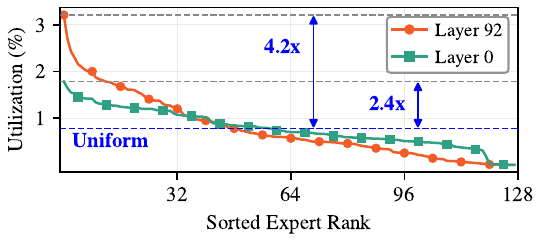}
    \caption{Expert utilization for Layers 0 and 92 of Qwen3-235B on ShareGPT. Some experts are used more often than others and the most used experts differ across layers.}
    \label{fig:expertimbalance}
\end{figure}

The problem of stragglers worsens because \moe models suffer from \textit{expert imbalance}, where a few experts receive a disproportionately large number of tokens as compared to the remaining experts. 
For example, for the Qwen3-235B model with 128 experts, if each expert was used uniformly, then the expert utilization rate would be ~0.78\%. However, as shown in Figure~\ref{fig:expertimbalance}, the most frequently used expert is used $4.2\times$ and $2.4\times$ more often than the idealized uniform utilization rate. Moreover, the utilization of each expert depends on the specific layer of the model and the distribution of the expert utilization for each layer is different from another.

\ignore{
If routing was uniform, each expert in a layer would receive a load of 100\%/$E_{total}$ tokens, where $E_{total}$ represents the number of experts in that layer. However, Figure~\ref{fig:background_figure}(a) shows that \moe routing is highly skewed in practice. Furthermore, the heavily utilized experts differ across layers. Even if all GPUs had identical performance, GPUs hosting the most utilized experts would take longer to finish, due to expert imbalance. We illustrate this in Figure~\ref{fig:background_figure}(b), here GPU~0 receives many more tokens than GPU-1 in Layer 0. Consequently, tokens cannot proceed to Layer 1 until GPU~0 finishes processing. Similarly, GPU~0 must wait for GPU~1 in Layer 1. As a result, even without performance variability between GPUs, expert imbalance introduces straggler GPUs at every layer. 
}


\ignore{
\label{sec:variability}
\subsection{Variability in GPU Systems}
Even perfectly balanced token routing would not eliminate stragglers. Modern GPUs exhibit significant performance \textit{ variability} due to many hardware and system-level factors. Process variation during manufacturing creates differences in clock frequencies and leakage characteristics across dies~\cite{sinha2022not, yoshida2022analyzing, tiwari2007recycle}. During runtime, dynamic voltage and frequency scaling (DVFS), which varies with physical placement and cooling mechanisms, causes GPUs in the same node to operate at different frequencies for the same workload. At the system level, OS scheduling, PCIe contention, and virtualization layers further contribute to performance differences across GPUs ~\cite{kurzynski2025lit,jeon2019analysis,chen2016baymax,chen2017prophet,xu2019characterization}.

\noindent We quantify these variability effects by measuring token generation speeds across several NVIDIA 8×L40 nodes, running Mistral-7B-Instruct-v0.3 with a batch size of 128 on vLLM. Figure~\ref{fig:background_figure}(c) shows that the 99th percentile token throughput varies by 16.2\% across GPUs. These differences in processing speed create stragglers in EP inference even when tokens are uniformly distributed. In Figure~\ref{fig:background_figure}(d) GPU 1 is a straggler due to performance variability, despite receiving a total token load identical to GPU 0. Effective expert mapping strategies must therefore account for both expert imbalance and performance variability
}

\newpage
\subsection{Prior Works On Addressing Stragglers}
\label{limitation}

Currently, there exists various prior works that address the straggler problem caused by expert imbalance. 

\noindent {\textbf{Training Strategies:}} These approaches use an auxiliary load balancing loss~\cite{fedus2022switch, lepikhin2020gshard} to penalize expert imbalance. However, this introduces a trade-off between model quality (primary training objective) and load balancing of experts. Setting the loss co-efficient too high forces artificial uniformity and degrades accuracy, while setting it too low causes the router to converge on a small subset of experts, starving the rest (called ``routing-collapse")~\cite{qiu2025demons, shazeer2017outrageously}. DeepSeek-V3~\cite{deepseekai2025deepseekv3technicalreport, wang2024auxiliary} mitigates this by introducing a bias term that adjusts routing without affecting training gradients, but even when trained with such techniques, models exhibit significant load imbalance at inference time~\cite{huang2024toward}. This is because token distributions during inference differ from that during training. This causes the router to produce skewed expert selections, despite how balanced the pre-training data was.

\noindent {\textbf{Inference Strategies:}}
Dynamic approaches, such as vLLM's Expert Parallel Load Balancer (EPLB)~\cite{kwon2023efficient, deepseekai2025deepseekv3technicalreport}, periodically re-distribute experts across GPUs based on observed load. However, this requires copying expert weights across GPUs, which introduces memory overheads. On the other hand, static approaches like MOETuner~\cite{go2025moetuner} and ExFlow~\cite{yao2024exploiting} compute expert mapping offline using profiled routing traces. 

\begin{figure}[htp]
    \centering
    \includegraphics[width=1\linewidth]{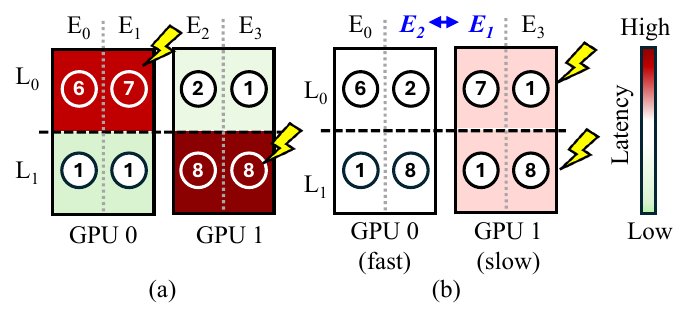}
    \caption{(a) Trivially allocating experts causes some GPUs to process significantly more tokens and become stragglers because some experts are used more often than others. (b) Prior works attempts to overcome this problem by redistributing experts to balance token loads across GPUs.}
    \label{fig:straggler}
\end{figure}

We illustrate this by using Figure~\ref{fig:straggler}. Initially, assume experts $E_0$, $E_1$ are placed on GPU-0 and $E_2$, $E_3$ are on GPU-1, as shown in Figure~\ref{fig:straggler}(a). But it turns out that the most frequently used experts for Layer-0 are $E_0$ and $E_1$, which causes GPU-0 to process a significantly larger number of tokens while GPU-1 remains idle, eventually making GPU-0 a straggler. Similarly, experts $E_2$ and $E_3$ are used more often for Layer-1, making GPU-1 the straggler for this layer. Prior methods redistribute the experts, moving $E_1$ to GPU-1 and $E_2$ to GPU-0 as shown in Figure~\ref{fig:straggler}(b), where the most frequently used experts are not co-located on the same GPU.

\subsection{Limitations of Prior Works}
\label{variabilitylimitation}

\begin{figure}[htp]
    \centering
    \includegraphics[width=1\linewidth]{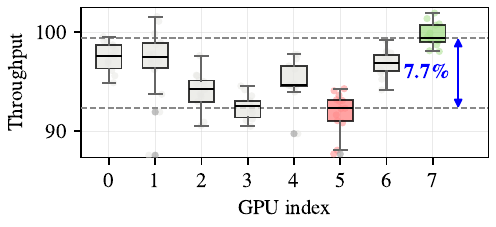}
     \caption{Throughput (tokens/s) for DeepSeek-R1-Distill-Qwen-7B on NVIDIA 8xL40s over a week shows that modern GPUs exhibit persistent performance variability.}
    \label{fig:perf}
\end{figure}

While more effective than trivial expert allocation methods, prior works still suffer from a critical drawback. They 
are agnostic of the variability in the underlying hardware. Our experiments show that even state-of-the-art Nvidia 8xL40 GPUs exhibit significant performance variability.  Figure ~\ref{fig:perf} shows the variation in throughput of the ShareGPT dataset for the DeepSeek-R1-Distill-Qwen-7B model served at batch size of 128 on a cluster with 8 GPUs over a week. We observe up to 7.7\% difference in the throughput between the best and worst GPUs.  
This aligns with observations from other prior work that modern GPUs exhibit significant performance variability~\cite{sinha2022not, jain2024palvariabilityawarepolicyscheduling}. Such heterogeneity directly impacts the performance of MoE models because even if we re-distributed the experts perfectly, some GPUs would continue to remain stragglers due to the inherent non-uniformity, such as GPU-1 in the illustration in Figure~\ref{fig:straggler}(b).

\begin{figure}[htp]
    \centering
    \includegraphics[width=1\linewidth]{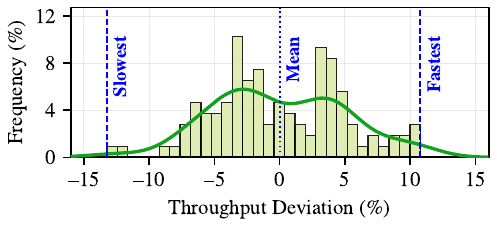}
    \caption{Distribution of throughput variation for DeepSeek-R1-Distill-Qwen-7B on 128 NVIDIA L40s. The fastest GPU profiled offers 27.7\% higher throughput than the slowest.}
    \label{fig:throughputdeviation}
\end{figure}

Figure~\ref{fig:throughputdeviation} shows the distribution of the variation in throughput across 16 different Nvidia 8xL40 GPUs for a total of 128 GPUs. The best GPU offers 10.8\% higher throughput compared to the average. In contrast, the worst GPU reduces the throughput by 13.23\% compared to the average. We present more characterization results in Appendix~\ref{app:variability}.

\textit{
\textbf{Our goal: GPU-variability-aware expert mapping} for MoE models to minimize stragglers and improve throughput.}

%% file: sections/3_Design.tex
\newtcolorbox{insightbox}{
  colback=blue!4,
  colframe=blue!55!black,
  boxrule=0.6pt,
  arc=2pt,
  left=6pt, right=6pt, top=4pt, bottom=4pt,
  fontupper=\itshape,
}

\section{Our Proposal: \papername}
\label{sec:design}
We propose \textit{\papername}, a framework for high-throughput inference using \moe models. \papername enables
GPU-variability-aware expert mapping to minimize the impact of stragglers. In this section, we discuss the challenges in enabling
\papername before describing how our design addresses them.

\subsection{Challenges in Designing \papername}
A natural strategy to enable GPU-variability-aware expert mapping would be to simply place the most heavily used experts on to the fastest GPUs. However, this approach is non-trivial due to two reasons.

\vspace{0.5em}
\noindent\textbf{Challenge-1: GPUs exhibit variability, but only limited}

\noindent While GPU performance variability indeed exists, the 
fastest GPU in a compute node is \textit{only modestly} faster than the slowest one. This limited gap constrains the extra token load that the fastest GPU can process. Naively placing too many frequently used experts on the fastest GPU exceeds this limit, eventually causing this GPU to become a straggler itself.

\vspace{0.5em}
\noindent\textbf{Challenge-2: Temporal nature of expert activations }

\begin{figure}[!h]
    \centering
    \includegraphics[width=0.95\linewidth]{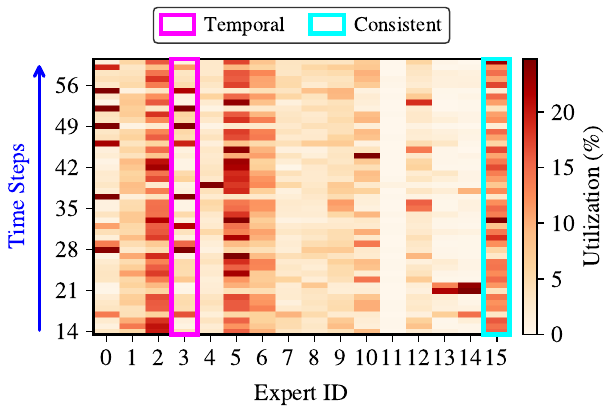}
    \caption{Per-step utilization for the 16 experts in Layer-43 of Llama-4 Scout. Consistent experts (2, 5, 15) are used in 85\% of timesteps, whereas temporal experts (0, 3, 10) are used in only 17\% of time steps but process 3x more tokens. }
    \label{fig:expert_pair_heatmaps}
\end{figure}

\noindent Most frequently used experts can often be grouped into two categories depending on their activations: (1) \textit{consistent experts} that are used in almost every time step and (2) \textit{temporal experts} that are often used together in a small fraction of time steps. In state-of-the-art LLM serving frameworks such as vLLM, a time step refers to the step where a token is generated for each request in a batch. If we were to simply monitor the average token load across $N$ steps to identify the most frequently used experts, we would only be able to detect the consistent experts while the temporal experts would go undetected because their \textit{average} individual token loads across $N$ steps is typically low. For example, Figure~\ref{fig:expert_pair_heatmaps} shows the activations for the 16 most frequently used experts in Layer~43 of the Llama-4 Scout model over many time steps.

Experts 2, 5, and 15 are the consistent experts used across 85\% of the time steps, whereas experts 0 and 3 are temporal experts that are heavily used together only in \textit{certain phases}. As temporal experts are often used simultaneously, they cause slowdown when placed on the same GPU. For instance, if experts 0 and 3 were placed on the same GPU, it would receive more than 50\% of the tokens in 10\% of time steps which is 2$\times$ the load (25\%) of a fully balanced 4-GPU system.

\subsection{Key Insights}
\label{subsec:challenges}

\begin{figure}[htp]
    \centering
    \includegraphics[width=1\linewidth]{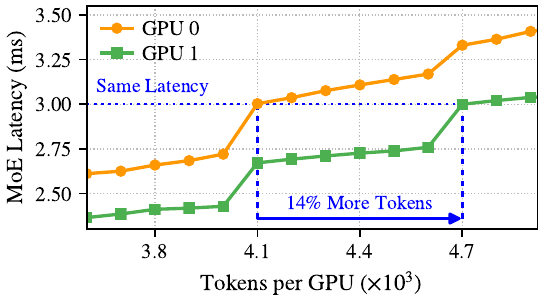}
    \caption{Latencies for \moe computation for Llama-4 Scout on an NVIDIA 8xL40 shows that latencies are identical even as GPU 1 receives 14\% more tokens than GPU 0.}
    \label{fig:complementary}
\end{figure}

\noindent\textbf{Insight-1: Balance end latencies, not token loads\\} Placing too many of the most frequently used experts on the fastest GPUs risks making those GPUs the new stragglers, whereas placing experts such that each GPU processes the same number of tokens does not mean they will all complete processing at the same time for each time-step. Our key insight is that rather than uniformly allocating token loads across GPUs (as done in prior works), we must place experts such that faster GPUs process \textit{only} proportionally more tokens than the slower ones based on the performance gap. For instance, Figure~\ref{fig:complementary} shows that GPU-1 (fastest) on an NVIDIA 8xL40 can process 14\% more tokens in the same amount of time as GPU-0 (slowest). Thus, experts must be placed by accounting for this relationship such that GPU-1 only hosts a subset of the most utilized experts such that it processes 14\% more tokens compared to GPU-0. 

\vspace{-0.1in}
\begin{insightbox}
Insight-1: We must place experts so that each GPU receives non-uniform token loads based on their performance and all GPUs finish processing simultaneously per step. 
\end{insightbox}

\begin{figure}[htp]
    \centering
    \includegraphics[width=1\linewidth]{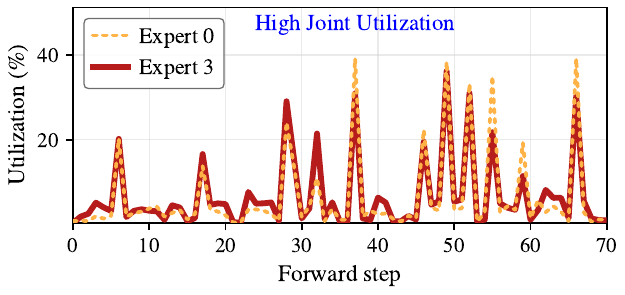}
    \caption{Experts 0 and 3 in Layer 43 of Llama-4-Scout are strongly correlated across engine steps ($r{=}0.88$). }
    \label{fig:expert_pair_correlations}
\end{figure}

\noindent\textbf{Insight-2: Optimize for both types of experts\\}
While most time steps are bottle-necked by consistent experts, a considerable fraction of them are impacted by the temporal ones. 
An ideal expert-to-GPU mapping policy 
must therefore reduce the slowdown caused by both types of experts. Placing consistent experts is relatively easier because the mapping policy can simply put these experts on separate GPUs. In contrast, placing temporal experts is non-trivial because the time steps in which they are utilized are not known in advance. 
However, we observe that the utilization of temporal experts is often non random; a subset of these experts are frequently used together. We refer to them as \textit{correlated temporal experts}. For instance, Figure~\ref{fig:expert_pair_correlations} shows that Experts 0 and 3 in Layer 43 of Llama-4 Scout have a Pearson's correlation coefficient of $0.88$ across time steps. The Pearson's correlation coefficient quantifies the similarity between two variables with values ranging between -1 to 1 and values closer to 1 denoting higher correlation~\cite{pearson1896}. By placing correlated temporal experts on different GPUs, their impact on overall slowdown can be minimized. 

\vspace{-0.1in}
\begin{insightbox}
Insight-2: Place both \textit{consistent} and \textit{correlated temporal experts} on different GPUs to minimize slowdown. 
\end{insightbox}
\subsection{Design Overview and Implementation}
\label{subsec:design-overview}

\begin{figure*}[t]
    \centering
    \includegraphics[width=\linewidth, trim={0 0 0 0}, clip]{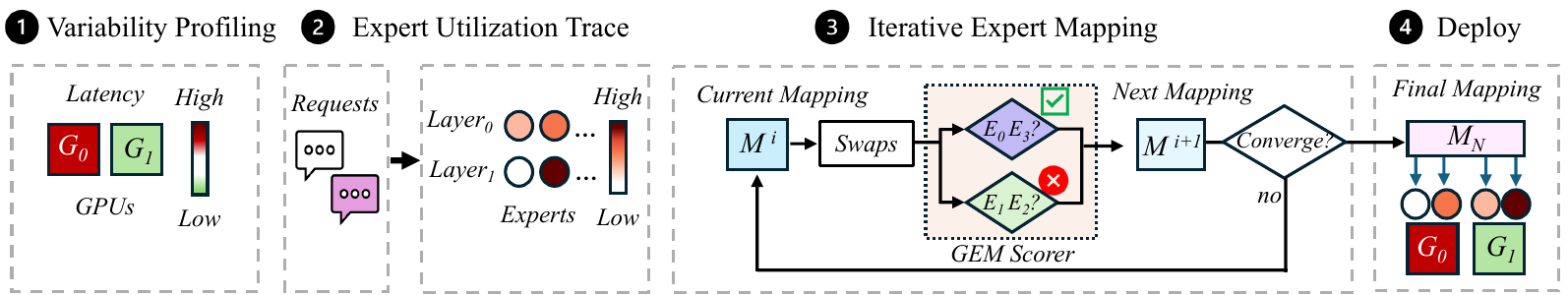}
    \caption{Overview of \papername. First variability is profiled across GPUs. Next, expert utilization patterns are captured to form a trace. Then, an iterative search refines expert mapping. Finally, the expert mapping is deployed on the system.}
    \label{fig:design}
\end{figure*}

\papername places experts by accounting for both expert utilization patterns of a task and the GPU performance variability. However, expert utilization depends on the task and is not known prior to inference. Moreover, the number of tokens processed by each GPU depends on the task and model. Thus, the impact of GPU variability on model performance for a given task is also unknown. To address this challenge, \papername employs four steps, outlined in Figure ~\ref{fig:design}. \circled{1} \papername assesses the expert utilization patterns of a task by counting the number of tokens routed to each expert during online inference. Initially, it places experts using vLLM's \cite{kwon2023efficient} default linear mapping to capture this information. \circled{2} Next, \papername profiles variability across GPUs offline by using a micro-benchmark to measure \moe-layer latency for different token loads. \circled{3} Then, \papername uses the expert utilization patterns and variability information to identify a better \textit{expert to GPU mapping} or an assignment of expert to GPUs to minimize stragglers. \circled{4}  Finally, \papername replaces the initial mapping (from Step-1) with this task-specific variability-aware mapping for every \moe layer during deployment. Such in-deployment expert swapping is similar to prior works like vLLM EPLB and has minimal overheads. Next, we discuss these steps in detail.

\newpage
\subsubsection{Step-1: Expert Utilization \Trace}

To optimally place experts on the GPUs, \papername must know how experts are typically used over time steps. To assess this, \papername counts the number of tokens each expert receives during online inference. Accessing this information is trivial because it is already created during inference by the \moe router when it computes the top-$k$ expert assignment for each token. \papername simply collects this information while serving requests
to form an \textit{expert utilization \trace} and then uses this trace to enable GPU-variability-aware expert mapping. 

\begin{figure}[htp]
      \centering
      \includegraphics[width=1\linewidth]{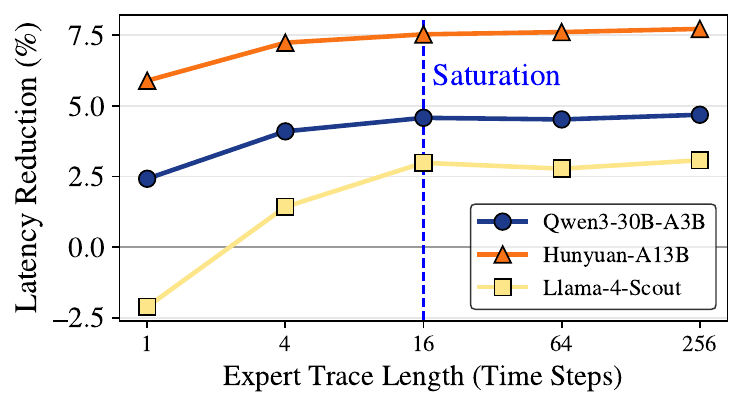}
      \caption{Latency reduction for Qwen3-30B-A3B, Hunyuan-A13B, and Llama-4-Scout on 500 unseen ShareGPT requests across \trace lengths. Even a short trace of 16 time steps suffice to build a robust variability-aware expert mapping.}
      \label{fig:tracewindowlength}
  \end{figure}

\vspace{0.05in}
\noindent\textbf{What is the right \trace length?} There exists a tradeoff between the quality of the expert utilization \trace and the time to deployment. A short \trace collected over a small number of time steps successfully captures the utilization patterns of the consistent experts but often fails to capture the utilization of the temporal experts because they are used in a small subset of time steps. In contrast, a longer \trace captures both experts but delays deployment and reduces the efficacy of \papername. Ideally, we want to find the shortest \trace length that captures the utilization patterns of both consistent and temporal experts. We conduct an experiment to confirm our hypothesis and identify the sweet operating point by sweeping the \trace length from 1 to 256, placing experts based on the \trace, and measuring the end-to-end latency reduction on 500 unseen requests from the ShareGPT~\cite{sharegpt} dataset.  Figure~\ref{fig:tracewindowlength} shows the relationship between the \trace length and end-to-end latency for three architecturally distinct \moe models: Qwen3-30B-A3B, Hunyuan-A13B, and Llama-4-Scout. From this study, we make two observations. \textit{First}, a single time step \trace does not fully capture the expert utilization patterns of temporal experts and therefore has high end-to-end latency, as expected. For Llama-4-Scout, this short \trace results in an expert mapping which has a 2.2 \% \textit{worse} end-to-end latency than vLLM's default linear expert mapping. This data highlights the importance of temporal experts to reducing latency. \textit{Second}, the performance saturates at 16 time steps for all three models. This indicates that expert utilization patterns are stable over time. If utilization had drifted, longer traces would continue to improve the mapping. Based on this study, \papername selects 16 as the default trace window length across models. 

\subsubsection{Step-2: Performance Variability Profiling} \hfill

\noindent \papername profiles the latency of the \moe-layers to capture the throughput variation across GPUs. For this, \papername uses a microbenchmark that launches an isolated \moe expert kernel with synthetic input batches at each target token count and records the average processing latency on each GPU. This produces a per-GPU token-vs-latency curve which is used to score expert mappings in subsequent steps.


\vspace{0.05in}
\noindent\textbf{Fast and accurate profiling:} 
Profiling at an extremely fine-granularity, such as each token, offers highly accurate profiles but requires more time during which the GPUs are unusable for actual inference tasks. For example, on a 4$\times$H200 NVL system, sweeping every token count from 1 to 16K with 500 kernel launches per count takes 12 hours for Llama-4-Scout~\cite{llama4}. Also, these profiles often become obsolete as environmental factors influencing dynamic voltage frequency scaling, such as cooling and power setup, change~\cite{sinha2022not, jain2024palvariabilityawarepolicyscheduling, patel2024polca}.
Performance variability also depends on the model, those with larger experts exhibit more variability than models with smaller experts~\cite{mixtral, llama4}. 
In contrast, too coarse fixed intervals of hundreds of tokens fails to capture the latency distribution accurately because latency does not increase smoothly with token count, but follows a \textit{staircase} pattern~\cite{gale2023megablocks, dao2022flashattention}. 
\papername exploits the insight that \moe networks group tokens into fixed-size batches called \textit{tiles}, typically multiples of 32 or 64 tokens depending on hardware and software configurations~\cite{gale2023megablocks, kwon2023efficient}
 and latency only jumps upon crossing tile boundaries. Figure~\ref{fig:complementary} illustrates this on Llama-4-Scout~\cite{llama4}, where latency increases sharply at a granularity of 512 tokens per GPU. Thus, the tile size offers the right granularity for both accurate and fast profiling.

\begin{figure}[!h]
    \centering
    \includegraphics[width=1\linewidth]{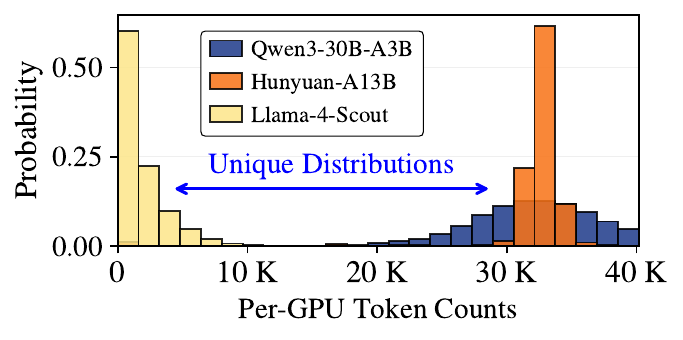}
    \caption{Per-GPU token counts for Qwen3-30B-A3B, Hunyuan-A13B, and Llama-4-Scout shows that each model receives a different number of tokens per GPU.}
    \label{fig:tokencountdist}
\end{figure}

\vspace{0.1in}
\noindent\textbf{Profiling Strategy:} \papername must account for variable token counts processed by each GPU depending on the model during profiling. This variation arises from differences in model architectures, like number of experts per layer and maximum batch size~\cite{fedus2022switch, mixtral, dai2024deepseekmoe}. For example, Figure~\ref{fig:tokencountdist} shows that token counts vary across the Llama-4-Scout~\cite{llama4}, Hunyuan-A13B~\cite{hunyuan}, and Qwen3-30B-A3B~\cite{qwen3} \moe models when deployed on a 4xH200. \papername therefore adapts the profiled token counts range for each model. For instance, \papername only profiles token counts between 1 and 10K tokens for Llama-4-Scout. Some models, however, have per-GPU token distributions that range up to tens of thousands of tokens, where sampling at even the granularity of tile size requires a large number of samples. We observe that at high token counts, the latency increase per tile shrinks to a small fraction of the total latency. In this range, \papername samples sparsely at intervals of thousands of tokens and reconstructs the full curve by linearly interpolating between the sampled points. Together, these policies make profiling fast: on a 4$\times$H200 NVL system, \papername generates a profile for Llama-4-Scout~\cite{llama4} within 3 minutes, sampling 62 token counts with 500 measurements each.
\subsubsection{Step-3: Variability-Aware Expert mapping} \hfill
\label{sec:algo}
\papername then performs the expert mapping step during which it decides on which GPU to place each expert.
For this assignment, the number of experts per GPU is determined by dividing the total number of experts by the number of GPUs, such that each GPU hosts the same number of experts. This ensures that the memory consumed by expert weights remains the same across GPUs and sufficient memory capacity is available to hold the KV caches (necessary to hold context information for fast inference) on each GPU. \papername must identify the optimal \textit{mapping} or a mapping of the experts on to the GPUs that minimizes the impact of stragglers by accounting for both expert utilization as well as GPU performance variability by exploiting the insights described earlier. However, identifying the optimal mapping is infeasible because the number of possible mappings grows combinatorial in the number of experts and GPUs.

To address this limitation, \papername employs a heuristic based iterative approach, where it selects an initial mapping, identifies its impact on end-to-end latency, uses this information to find an alternate mapping, and updates the current mapping to this alternate candidate for the subsequent iteration. The process is repeated until the mapping does not improve any further. However, the performance of each iteration depends on the initial mapping that \papername started with. To avoid convergence on sub-optimal mappings, \papername repeats this iterative approach starting with multiple initial mappings, each called a restart. Finally, \papername selects the best expert mapping across all restarts. Our experiments show that about 30 restarts are sufficient for achieving good performance and we use this as the default in our design. Nonetheless, it may be adjusted in the future as models and GPUs evolve. 

\begin{figure}[htp]
    \centering
    \includegraphics[width=\linewidth]{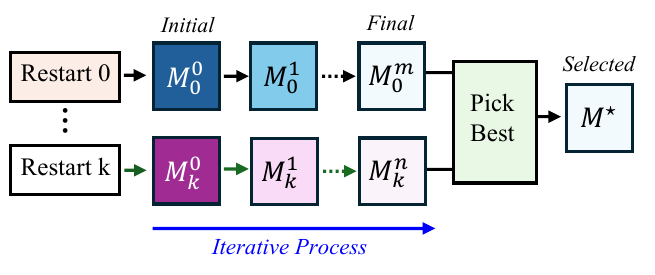}
    \caption{\papername's expert mapping starts with $K$ initial mappings, each iteratively refined until latency saturates, and then selects the final mapping with lowest latency.}
    \label{fig:mapping_algo}
\end{figure}

Figure~\ref{fig:mapping_algo} illustrates this selection process, where $M_i^j$ refers to the expert to GPU mapping at the beginning of the $j^{th}$ iteration of the $i^{th}$ restart.
\papername generates the initial expert mapping $M_i^0$ for the $i^{th}$ restart.
From each $M_i^0$, \papername runs an iterative refinement procedure yielding a sequence of subsequent mappings ($M_i^1, M_i^2, \ldots, M_i^m$) which terminates when the latency does not improve further.
\papername then selects the mapping $M^\star$ with the lowest latency among the $K$ final mappings ($M_0^m, \ldots, M_k^n$) for deployment. 

\vspace{0.4em}\noindent\textbf{Scoring Expert Mappings:} To assess a mapping, \papername replays the trace from Step-1 in software and predicts the latency of each step (to account for expert utilization) by using the variability profiles from Step-2 (to account for non-uniformity in hardware). Recollect that faster GPUs should process proportionally more tokens than slower ones so that all GPUs finish at the same time (\textit{Insight-1}). So, the scoring function, $S()$, captures latency rather than actual token counts. \papername also ensures that consistent and correlated temporal experts are placed on different GPUs to avoid slowdown (\textit{Insight-2}) by scoring each step separately to appropriately place both types of experts. Given a mapping $M$, trace $T$, with per-GPU latency cost $C_g(\cdot)$, \papername scores the mapping as shown in Equation~\ref{eq:score}, where $n_g(M, t)$ is the number of tokens routed to GPU $g$ at step $t$ for a given mapping $M$. The $\max()$ captures the straggler GPU at each step and the outer summation accumulates the impact of stragglers across the entire trace. 

\begin{equation}
S(M) \;=\; \sum_{t \in T} \; \max_{g} (\; C_g\!\left(n_g(M, t)\right))
\label{eq:score}
\end{equation}

\begin{figure}[htp]
      \centering
      \includegraphics[width=1.\linewidth]{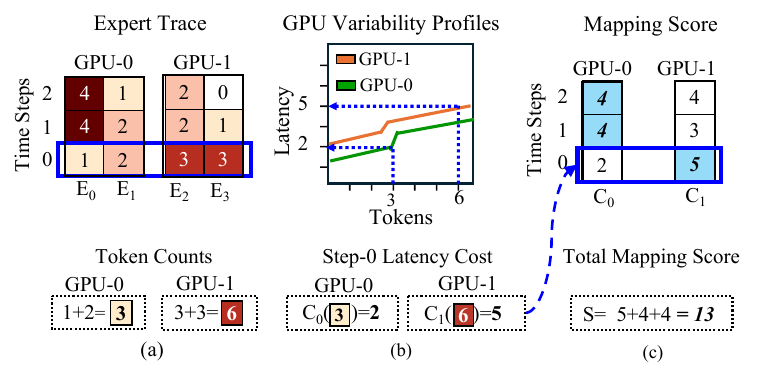}
      \caption{ (a) Example of an expert trace used to determine the token count of each time step. (b) which is then used to compute the latency cost per GPU by consulting the variability profiles. (c) The mapping score is computed by summing the cost of the straggler GPUs over all time steps in the trace.}
      \label{fig:objective}
  \end{figure}

We explain this scoring process using the illustration in 
Figure~\ref{fig:objective}. Consider the time step 0 of the expert trace depicted in Figure~\ref{fig:objective}(a), where experts $E_0, E_1, E_2, E_3$ each receive 1, 2, 3, 3 tokens respectively. We compute per-GPU token counts for this step as $1+2=3$ and $3+3=6$ for GPU-0 and GPU-1 respectively. Then, the variability profiles of GPU-0 and GPU-1 are consulted to determine latencies to process these many tokens in each GPU, as shown in Figure~\ref{fig:objective}(b). So, for step 0, the latency cost of processing the tokens is computed as $C_0(3)=2$ for GPU-0 and $C_1(6)=5$ for GPU-1. Finally, the score for the mapping is computed as the sum of straggler latencies over all time steps. Figure~\ref{fig:objective}(c) shows that for time steps 0, 1, and 2, GPU-1, GPU-0, and GPU-0 are the straggler GPUs respectively with latencies 5,4, and 4 respectively. These latencies are summed as $5+4+4=13$ to compute the score for this expert mapping. 

\ignore{\vspace{0.4em}\noindent\textbf{Search Algorithm.} 

Algorithm~\ref{alg:search} outlines the iterative search process. \papername forms each initial expert mapping by sorting experts in reverse order by average expert utilization and placing each one on the GPU where it adds the least to the current score. The first restart forms an initial mapping with the expert utilization order as-is, whereas subsequent restarts add a noise component of 20\% to ensure each initial mapping is unique. \papername then evaluates potential swaps between pairs of experts on different GPUs and applies the swap that most reduces the current score. The search terminates once the best swap improves latency by less than 0.1\%. Figure~\ref{fig:search} shows that a stable convergence is reached in 18, 16, and 6 swaps for Qwen3-30B-A3B, Hunyuan-A13B, and Llama-4-Scout respectively. Finally, the best mapping from all k restarts is selected to place experts. 
}

\noindent\textbf{Finding Alternate Mappings For Iterative Search:} The objective of \papername's iterative search algorithm, outlined in Algorithm~\ref{alg:search}, is to take a mapping as an input and find an alternative mapping that outperforms it. 
The search process is based on a simple intuition: if $M_i^j$ is sub-optimal, then there exists two experts that are poorly placed in the current mapping. \papername therefore tries to find this pair of experts and swaps them to find $M_i^{j+1}$ such that $S(M_i^{j+1})<S(M_i^{j})$. Of all such expert pairs, \papername selects the pair with minimal $S(M_i^{j+1})$. This procedure is repeated until no better mapping is found, terminating at $M_i^m$. A key challenge of running this type of search is that the quality of the final mapping depends on the initial mapping. Thus, to improve the quality of the final mappings and optimize the search process, \papername therefore starts with carefully chosen values of $M_i^0$. This initial mapping starts with all GPUs empty. Then, \papername sorts all experts from the most to the least utilized and places each expert on the GPU that minimizes $S(M_i^0)$. This heuristic operates based on the intuition that the most used experts have the maximum impact on latency and therefore, they must be placed first. But this starting point may result in a sub-optimal final mapping $M_i^m$ due to the heuristic nature of the search. So, \papername operates with $k$ initial mappings $M_0^0, \ldots, M_k^0$. To ensure that these mappings are unique, \papername adds a 20\% noise component to the utilization metric used to sort experts. Our experiments show that this search algorithm converges in under 18 expert swaps for all models evaluated, meaning that $M_i^0$ is already a good initial mapping, and the search algorithm just refines it. More detailed description of our algorithms are presented in Appendix~\ref{app:algorithm}.


\begin{algorithm}[ht]
\caption{\textsc{\papername's Search Algorithm}}
\label{alg:search}
\begin{algorithmic}[1]
\Require utilizations $u$
\Ensure expert-to-GPU mapping $M_m$
\State $M_0 \gets \emptyset$ \textcolor{blue}{\textit{   // Initial mapping: place heaviest experts first}}
\For{$e$ in experts sorted by $u{+}0.2\!\cdot\!\text{noise}$}
    \State place $e$ on GPU minimizing increase in $S(M_0)$
\EndFor
\Repeat \textcolor{blue}{\textit{   // Refine via best cross-GPU swap}}
    \State $s \gets S(M_i)$
    \State $(e_a, e_b) \gets$ swap expert pair that most reduces $S(M_i)$
    \State $M_{i+1} \gets \textsc{Swap}(M_i, e_a, e_b)$
\Until{$1 - S(M_i)/s < 0.001$}
\State \Return $M_m$
\end{algorithmic}
\end{algorithm}


\subsubsection{Step-4: Deployment and Serving} 

\noindent Once Step-3 produces an expert mapping for each \moe layer, \papername loads each expert's weights onto its assigned GPU at model load time, thereby replacing the initial linear mapping with the task-specific GPU variability-aware mapping. The \moe router continues to dispatch each token to its top-k experts as before, but the experts now reside on GPUs chosen by \papername's variability-aware search. 

\vspace{0.05in}
\noindent\textbf{Time to deployment:} \papername is very fast. For instance, variability profiling for Llama-4-Scout takes 2.13 minutes and mapping takes just 8.8 seconds. These costs are negligible and the benefits of \papername far outweigh this 
latency in long-running deployment scenarios that serve requests for hours or days~\cite{wang2024burstgpt, goel2025sageserve}.

%% file: sections/4_Methodology.tex
\clearpage

\section{Evaluation Methodology}
\label{sec:methodology}

We discuss the methodology used to evaluate \papername.

\subsection{Models}
We evaluate \papername on five state-of-the-art \moe models spanning the Mixtral~\cite{mixtral}, Hunyuan~\cite{hunyuan}, Qwen3~\cite{qwen3}, and Llama-4~\cite{llama4} families comprising 8 to 128 routed experts per layer and 30B to 141B total parameters, as shown in Table~\ref{tab:models}. Such a comprehensive evaluation across diverse \moe architectures shows \papername's widespread applicability.

\begin{table}[ht]
\setlength{\abovecaptionskip}{0pt}
\setlength{\belowcaptionskip}{4pt}
\centering
\caption{\moe models used to evaluate \papername.}
\label{tab:models}
\small
\renewcommand{\arraystretch}{1.3}
\setlength{\tabcolsep}{8pt}
\begin{tabular}{|l|c|c|c|}
\hline
\textbf{Model} & \textbf{Layers} & \textbf{Experts/Layer} & \textbf{Params} \\
\hline
Mixtral-8x7B    & 32 & 8   & 47B  \\
\hline
Mixtral-8x22B   & 56 & 8   & 141B \\
\hline
Llama-4-Scout   & 48 & 16  & 109B \\
\hline
Hunyuan-A13B    & 32 & 64  & 80B  \\
\hline
Qwen3-30B-A3B   & 48 & 128 & 30B  \\
\hline
\end{tabular}
\end{table}

\subsection{Experimental Setup}
We conduct our experiments on a compute node with 4 NVIDIA H200 GPUs (141~GB HBM3 each) and a dual-socket AMD EPYC 9354 32-core CPU. All GPUs are connected through a full-mesh NVLink fabric (900~GB/s per GPU). We implement \papername in vLLM~\cite{kwon2023efficient}, a state-of-the-art LLM inference framework. We deploy \moe models with expert and tensor parallelism ranks of 4 to reduce memory requirements and enable large batches.

\begin{figure}[htp]
    \centering
\includegraphics[width=1\linewidth]{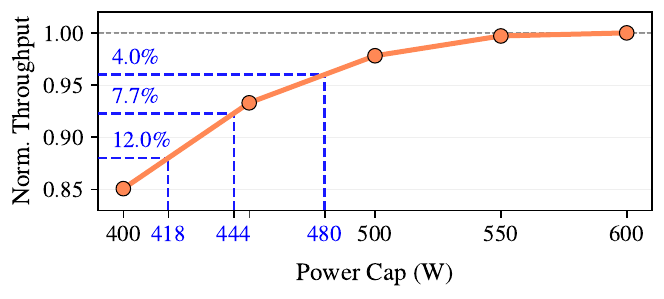}
    \caption{We profile power caps vs throughput to select per-GPU caps that emulate a desired variability setup.}
    \label{fig:powercaps}
\end{figure}

\noindent \textbf{Emulating Performance Variability:} As we only have access to a limited number of GPUs (only 4), we scale their performance (throughput) to emulate the performance variability observed at large server scale. This approach is based on prior works~\cite{krzywaniak2022gpu, patel2024polca, chung2024perseus}.
We emulate three different GPU setups corresponding to three unique performance variability characteristics. We derive these emulated setups from the throughput variation observed across 128 NVIDIA L40 GPUs during our characterization experiments. The studies show that the slowest GPU is 12\% slower than the average GPUs, whereas the fastest GPU is 11\% faster. The \textit{low-variability} setup assumes all 4 GPUs in the cluster have similar performance, corresponding to the mean GPU performance from the characterization. The \textit{moderate-variability} setup represents the average variability differences observed in our characterization studies. To obtain this, we average variation across 1000 Monte Carlo samples~\cite{metropolis1949monte} of size four from the throughput distribution.
The \textit{high variability} setup assumes a single straggler GPU which is 12\% slower than others, consistent with the slowest GPU from our characterization study. 
We calibrate per-GPU power caps to match the requirements of each variability setup. Figure~\ref{fig:powercaps} shows this calibration for for Llama-4-Scout. We invert this curve to pick power caps that produce each normalized throughput, listed in Table~\ref{tab:powercaps}.

\begin{table}[ht]
\setlength{\abovecaptionskip}{0pt}
\setlength{\belowcaptionskip}{0pt}
\centering
\caption{Per-GPU power caps (W) for each variability profile.}
\label{tab:powercaps}
\small
\renewcommand{\arraystretch}{1.3}
\setlength{\tabcolsep}{8pt}
\begin{tabular}{|l|c|c|c|c|}
\hline
\textbf{Variability} & \textbf{GPU 0} & \textbf{GPU 1} & \textbf{GPU 2} & \textbf{GPU 3} \\
\hline
\textit{High}  & 418 & 600 & 600 & 600 \\
\hline
\textit{Moderate} & 418 & 444 & 480 & 600 \\
\hline
\textit{Low} & 600 & 600 & 600 & 600 \\
\hline
\end{tabular}
\end{table}

\subsection{Baselines}
We compare \papername against two expert-mapping strategies natively supported in vLLM.

\begin{enumerate}[leftmargin=0cm,itemindent=.5cm,labelwidth=\itemindent,labelsep=0cm,align=left, itemsep=0.1 cm, listparindent=0cm]
    \item \textit{Linear (vLLM default)} splits experts into equally sized contiguous groups by index, such that each expert $i$ on an $N$ GPU system is placed on GPU  $\lfloor i/N \rfloor$
    \item \textit{EPLB (vLLM's expert-parallel load balancer)} sums per-GPU token counts across time steps and periodically redistributes experts to balance load.
\end{enumerate}

\subsection{Datasets}
We evaluate \papername on 4K requests from two datasets representative of LLM workloads.

\begin{enumerate}[leftmargin=0cm,itemindent=.5cm,labelwidth=\itemindent,labelsep=0cm,align=left, itemsep=0.1 cm, listparindent=0cm]
    \item \textit{ShareGPT~\cite{sharegpt}} is a collection of real-world conversations sourced from a live inference server. 
    \item \textit{CodeContests~\cite{codecontests}}, sourced from Codeforces, is a competitive programming dataset representing technical workloads.

\end{enumerate}




\subsection{Figure-of-Merit}
We use two metrics. First, \textit{end-to-end latency}, defined in Equation \eqref{eq:e2e_latency}, which is the elapsed time from the arrival of the request $t_{\text{arrival}}$ to the emission of the final token $t_{\text{finish}}$:
\begin{equation}
    t_{\text{e2e}} = t_{\text{finish}} - t_{\text{arrival}},
    \label{eq:e2e_latency}
\end{equation}

\noindent Second, we report the \textit{90th-percentile time-per-output-token} (p90 TPOT), defined in Equation~\eqref{eq:p90_tpot}:
\begin{equation}
    \text{p90 TPOT} = Q_{0.90}\!\left(\Delta t\right),
    \label{eq:p90_tpot}
\end{equation}
where $\Delta t$ denotes inter-token latencies (time between consecutive output tokens) across all decoded tokens in the evaluation set, and $Q_{0.90}(\cdot)$ is the empirical 90th percentile.  This metric quantifies the latency of the slowest decode steps, which affect users' perceived streaming smoothness. 

%% file: sections/5_Results.tex
\newpage
\section{Results}
\label{sec:results}

\begin{figure*}[!t]
  \centering
  \includegraphics[width=\linewidth]{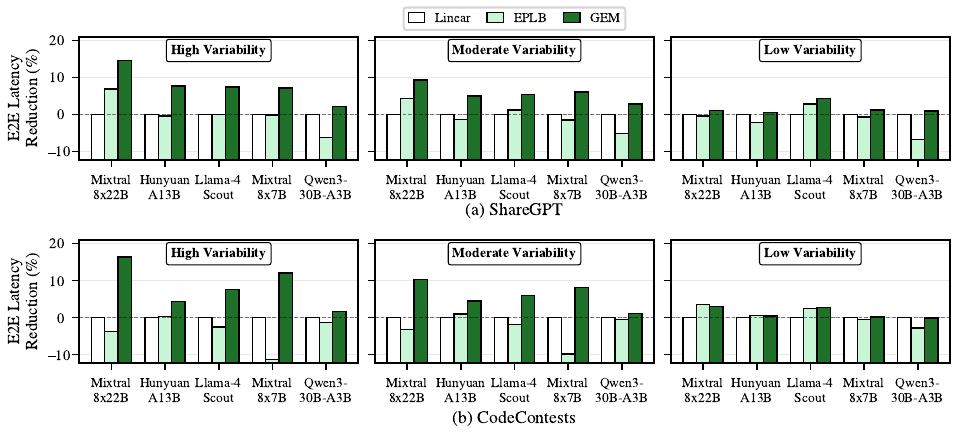}
  \caption{End-to-end latency reduction relative to linear mapping across all five \moe models (higher is better) on (a) ShareGPT and (b) CodeContests for three different GPU variability setups (high, medium, low).}
  \label{fig:results-e2e}
\end{figure*}


In this section, we evaluate \papername against baselines for two datasets, compare expert mappings against baselines, and quantify the cost of \papername's variability-profiling.

\subsection{End-to-End Latency} 
\label{sec:results-throughput} 
\papername outperforms both baselines across all configurations, as shown in Figure~\ref{fig:results-e2e}. As expected, the efficacy of \papername depends on the GPU variability. The end-to-end latency for the high-variability setup is reduced by {7.9\%} on average and by up to {16.5\%} in the best-case. For the  moderate-variability setup, it reduces by {5.9\%} on average and by up to {10.4\%}. Lastly, for the low-variability setup, it reduces by {1.5\%} on average and by up to {4.5\%}. We observe two key trends. \textit{First}, models with a few large experts concentrate utilization on a small number of GPUs and benefit the most from variability-aware mapping. For example, Mixtral-8x22B benefits the most on the high and moderate variability setups (by up to {15.0\%} and {9.9\%}). In contrast, Qwen3-30B-A3B's near-uniform routing across 128 small experts benefits the least (around {1.5\%}). \textit{Second}, even when GPU variability is low, \papername still offers latency reduction by appropriately placing correlated temporal experts. For instance, Llama-4-Scout has the most improvement ({3.5\%}) because it is rich in temporal experts while Qwen3-30B-A3B has the least ({0.4\%}).

\subsection{Tail-Latency (p90 TPOT)} 
\label{sec:results-p90-tpot} 
\begin{figure}[!h]
      \centering
      \includegraphics[width=0.98\linewidth]{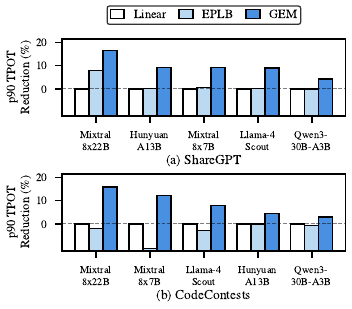}
      \caption{p90 TPOT latency reduction over linear mapping on the high-variability profile, for (a) ShareGPT and (b) CodeContests. Bars are
  normalized against linear mapping.}
      \label{fig:p90-tpot-highvar}
  \end{figure}

\papername's gains are most pronounced on tail latency which is critical to perceived responsiveness. Figure~\ref{fig:p90-tpot-highvar} shows that \papername reduces p90 TPOT by {9.1\%} on average and by up to {16.9\%} in the best case, exceeding the mean-latency improvements. This is because per-step latency is bounded by the \textit{straggler} GPU, and the p90 tail is dominated by steps where heavily used or correlated experts are placed on a slow GPU. \papername reduces tail latency by placing experts to balance per-GPU latencies (\textit{Insight-1}) and separating correlated temporal and consistent experts across GPUs (\textit{Insight-2}). We present tail latency results for all variability setups in Appendix~\ref{appendix:tail-latency}.

\begin{figure}[!hpt]
    \centering
    \includegraphics[width=\linewidth]{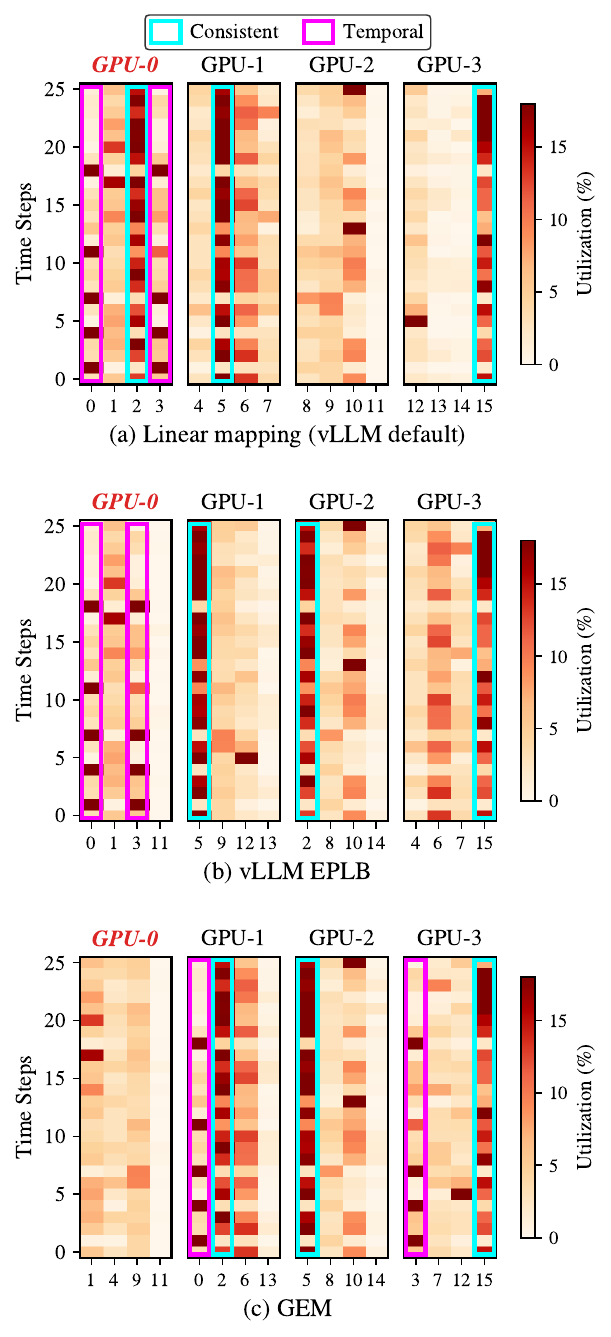}
    \caption{Comparison of expert mapping strategies for Layer 43 of Llama-4-Scout on the high-variability setup. (a) vLLM default \textit{linear} policy places experts based on indices, resulting in consistent and temporal experts being placed on GPU-0 (slowest). (b) The EPLB policy improves the allocation of the consistent experts but the temporal experts still remain on GPU-0. (c) \papername's expert mapping places both temporal and consistent experts on separate GPUs and avoids placing highly utilized experts on GPU-0.}
    \label{fig:mapping_heatmaps}
\end{figure}

\newpage

\subsection{Comparison of Expert Mapping Policies}
We compare the expert mapping of \papername against the two baselines using a four GPU setup, where GPU-0 is the slowest. For illustration, we only show the utilization of 16 experts of Layer 43 of Llama-4-Scout over 25 time steps.
The vLLM default, linear expert mapping, allocates four experts to each GPU in a linear fashion based on expert indices, such that experts 0 to 3 are placed on GPU-0, experts 4 to 7 are placed on GPU-1, and so on, as shown as Figure~\ref{fig:mapping_heatmaps}(a). This is sub-optimal because this expert-to-GPU allocation results in placing consistent expert 2 and temporal experts 0 and 3 on the slowest GPU-0. vLLM's expert-parallel load balancer, EPLB, adjusts this allocation and places the consistent expert 2 on GPU-1, as shown in Figure~\ref{fig:mapping_heatmaps}(b). As a result, EPLB reduces end-to-end latency by 1.0\% compared to the default linear mapping. However, temporal experts 0 and 3 still remain on GPU-0. In contrast, \papername optimizes the GPU allocation for both consistent and temporal experts. As shown in Figure~\ref{fig:mapping_heatmaps}(c), \papername ensures that the consistent experts 2, 5, and 15 as well as temporal experts 0 and 3 are allocated different GPUs and are not allocated GPU-0. It tries to minimize the overall utilization of GPU-0 compared to the other GPUs. As expected, such a GPU-variability-aware expert mapping enables \papername to reduce end-to-end latency by 6.2\% and 7.1\% compared to the EPLB and linear mapping policies respectively.

\ignore{We explain differences in end-to-end latency by comparing \papername's expert mapping to those of its baselines, EPLB and linear expert mappings. Figure~\ref{fig:mapping_heatmaps} visualizes the utilization and mapping of experts for Layer~43 of Llama-4 Scout over 20 time steps. The two highlighted columns track two correlated temporal experts as they are placed by each of the strategies. \textit{Linear} assigns contiguous experts to the same GPU, which places highly utilized \textit{consistent} and  \textit{correlated-temporal} experts on the slowest GPU. \textit{EPLB} places experts to balance aggregate per-GPU token counts across steps. However, EPLB still places the correlated temporal experts on the slowest GPU. By contrast, \papername places both highly utilized \textit{consistent} experts and correlated \textit{temporal} on separate GPUs, and avoids placing highly utilized experts on the slowest GPU. This mapping minimizes slowdown, and is responsible for the end to end latency reduction observed for \papername across models and datasets.                  }

\subsection{Variability Profiling Latency}

\begin{figure}[!h]
    \centering
\includegraphics[width=1\linewidth]{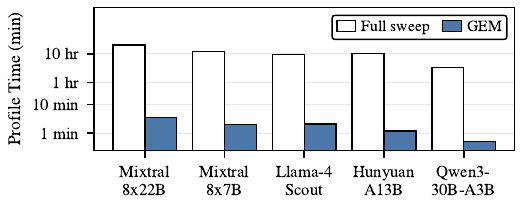}
\caption{\papername reduces the latency to build per-GPU token-vs-latency curves as compared to a full sweep which profiles every token count from 1 to 16 K.}
\label{fig:profile-time}
\end{figure}

Figure~\ref{fig:profile-time} shows that \papername completes variability profiling for all five \moe models within 0.5 to 3.6 minutes by sampling only at token counts where latency actually changes for each model. This approach significantly reduces the latency compared to a fine-grained sampling method which sweeps token counts from 1 to 16K and takes 3.4 to 20.5 hours. This results in a 265 to 515$\times$ reduction in the offline profiling cost. The savings are largest for models with wider FFNs, such as Mixtral-8x22B~\cite{mixtral}, Mixtral-8x7B~\cite{mixtral}, and Llama-4-Scout~\cite{llama4}, where each kernel launch is expensive. These latency savings result in lower system downtimes and reduce time to deploy \papername in real inference serving frameworks. 

\section{Performance Variability at Scale}
  \label{subsec:variability-scalability}

 \noindent The variability quantified in Section~\ref{variabilitylimitation} is a problem that is expected to worsen for large-scale GPU
  deployments as both models and systems scale. This is because the performance gap between the slowest GPU and the fastest scales as more GPUs are aggregated into a single deployment. To model the effects, we use the throughput
  distributions of the 128 NVIDIA L40 GPUs we characterized. For each system size $N$ we sample from the distribution using 10{,}000 Monte Carlo
  samples~\cite{metropolis1949monte} and compute the throughput gap between the slowest sampled GPU and the fastest.

  \begin{figure}[!h]
      \centering
      \includegraphics[width=0.9\linewidth]{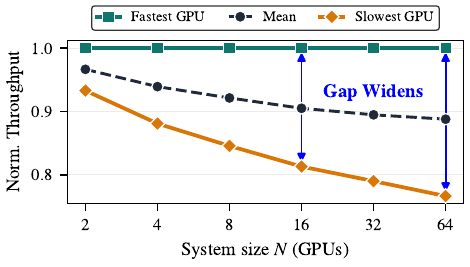}
      \caption{Expected per-GPU throughput across $N$ GPUs, normalized to the fastest GPU, shows that the effect of variability grows with system size $N$.}
      \label{fig:nscaling_variability}
  \end{figure}

\noindent Figure~\ref{fig:nscaling_variability} shows that the expected slowest-to-fastest throughput gap grows monotonically with $N$, increasing from
  {11.9\%} at $N{=}4$ (the scale of our evaluation setup) to {23.4\%} at $N{=}64$. This trend shows the importance of
  variability-aware expert mapping at scale, where straggler effects dominate.

%% file: sections/6_Discussion.tex

%% file: sections/7_Related.tex
\section{Related Work}
\label{sec:related}

In this section, we discuss related work and compare or contrast as appropriate. 

\vspace{0.05in}
\noindent \textbf{Reducing communication overheads:}
Various prior works focus on reducing the impact of communication between GPUs in expert parallel settings. 
MoETuner~\cite{go2025moetuner} places experts correlated across \textit{adjacent layers} on the same GPU. However, this yields limited latency benefits, as each token must return to its origin GPU (where its KV-cache resides) before being re-dispatched to the next layer's experts. 
ExFlow overcomes this by replicating the KV-cache of each request across all GPUs, but introduces memory overheads that limits batch sizes. Occult~\cite{luo2025occult} places experts which are likely to be used together on the same GPU, but this approach concentrates heavy experts, exacerbating expert imbalance. 
Speculative-MOE~\cite{li2025speculativemoe} reduces latency by dispatching tokens earlier to GPUs which are predicted to be used in advance of routing decisions. These approaches are orthogonal to \papername because they all overlook the non-uniformity of the underlying GPU hardware. The methods can be combined with \papername for even higher benefits.

\ignore{These approaches overlook the non-uniformity in GPU hardware and frequently place experts in ways that contradict with \papername's key insights. }

\ignore{These approaches are orthogonal to \papername because they all overlook the non-uniformity of the underlying GPU hardware. The methods can be combined with \papername for even higher benefits.
}


\vspace{0.05in}
\noindent \textbf{Dynamic expert provisioning:} Several prior works handle expert imbalance by replicating high utility experts. These methods primarily differ in how they decide which experts to replicate. EPLB~\cite{deepseekai2025deepseekv3technicalreport} uses recent expert utilization data to predict which experts are likely to be used.  Lina~\cite{li2023lina} uses prior layers' expert utilization to predict future layers' experts. Libra~\cite{yang2026libra} directly runs hidden states through future layers' MoE routers. HarMoEny~\cite{harmoeny2025} waits for exact routing results to replicate experts. In all cases, the parameter weights of replicated experts compete with KV-caches for GPU memory, constraining context lengths \cite{craft2025}. These works are agnostic of performance variability across GPUs. 

\vspace{0.05in}
\noindent \textbf{Training-Time MoE optimization:}
A separate line of work introduces mechanisms during training to encourage balanced expert activation. For instance Switch Transformer \cite{fedus2022switch} and GShard \cite{lepikhin2020gshard} use expert-imbalance losses during training. However, this interferes with the model's training objective and degrades accuracy \cite{wang2024auxiliary, zhou2022mixtureofexpertsexpertchoicerouting}. By contrast, Loss-Free Balancing \cite{wang2024auxiliary} adds routing biases to steer tokens toward underused experts, but this weakens expert specialization \cite{deepseekai2025deepseekv3technicalreport}. Expert choice \cite{zhou2022mixtureofexpertsexpertchoicerouting} balances loads by training experts to select top-k tokens (instead of tokens selecting top-k experts). However, this makes generation non-deterministic since the token composition of a batch changes which experts each token uses. Moreover, these methods do not address inference-time hardware heterogeneity, where identical expert loads produce different latencies across GPUs.


\ignore{
\subsection{Summary}

Prior work improves MoE inference efficiency along four axes: reducing inter-GPU communication, dynamically provisioning hot experts, balancing routing during training, and co-designing the model with hardware. All four assume that GPUs in the cluster run at identical speeds, and the inference-time approaches balance only the aggregate token load across GPUs. Neither assumption holds in practice. GPUs in the same node exhibit persistent performance variability (Section~\ref{variabilitylimitation}), so balancing token counts equally across GPUs leaves the slowest GPU as the straggler at every layer. Furthermore, a few experts are activated together only in short bursts during certain phases of a request, and their utilization is invisible to aggregate statistics. \papername addresses both gaps. It profiles each GPU's throughput and assigns tokens in proportion to its measured speed, so all GPUs reach the synchronization barrier at the same time. It scores candidate mappings at per-step granularity over the full trace, so it separates correlated temporal experts across GPUs even when their average load is low.}

%% file: sections/8_Conclusion.tex
\section{Conclusion}
\label{sec:conclusion}
Mixture-of-Experts (MoE) models reduce the per-token computational cost of LLM inference. However, their end-to-end latency is increased by \textit{straggler} GPUs at every layer-synchronization barrier which are caused by differences in expert utilization across GPUs and hardware performance variability. Prior works place experts to balance token counts processed by each GPU but remain unaware of the underlying hardware variability. Therefore, this mapping causes the slowest GPU to remain a persistent straggler even if each GPU receives an equal number of tokens. In this paper, we propose \papername which is to our knowledge the first hardware variability aware expert mapping framework for \moe inference. \papername exploits two key insights: (1) that latencies, not token counts should be balanced across GPUs, and (2) that heavily used experts (both \textit{consistent} and \textit{correlated-temporal}) must be separated across GPUs. \papername applies these insights to construct a four-step pipeline that captures an expert utilization trace, profiles \moe-layers at different token sizes, runs an iterative search to find a task-specific, GPU-variability-aware expert mapping, and deploys the resulting mapping during inference. Across five state-of-the-art \moe models and two representative workloads \papername improves end-to-end latency by 7.9\% on average and up to 16.5\% over baseline expert mapping strategies.

%% file: sections/9_Acknowledgements.tex
\section{Acknowledgments}
\label{sec:acknowledgements}
We thank Sujay Sanghavi for some of the technical discussions during the course of this project and acknowledge the Texas Advanced Computing Center (TACC) and the Center for Generative AI (CGAI) at UT Austin for providing computational resources that helped develop the research results reported in this paper. We thank the generous support from the Cockrell School of Engineering and the Amazon AI PhD Fellowship Program through the Amazon Science Hub at UT Austin, including the AWS cloud credits provided through the fellowship, which enabled the large-scale experiments conducted in this work. Poulami Das acknowledges the generous support through the AMD endowment at UT Austin.

%% file: sections/A0_VariabilityData.tex
\newpage

\begin{figure*}[htp]
    \centering
    \includegraphics[width=\textwidth]{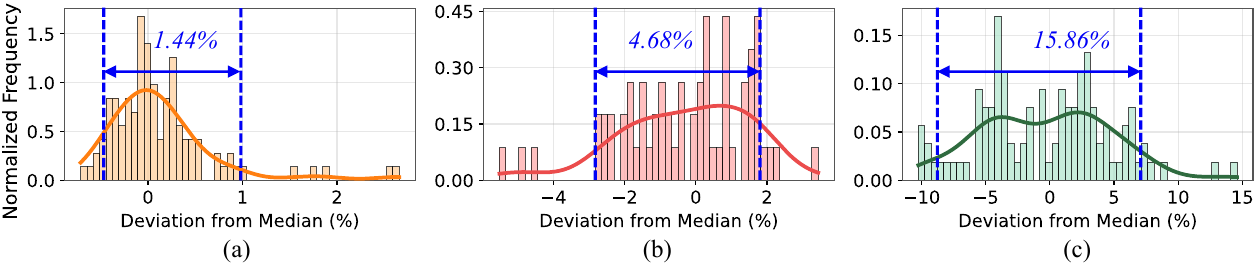}
    \caption{Aggregated TPOT variability across nodes for (a) DeepSeek-R1-Qwen-7B on Amazon Trainium (BS=32), (b) Qwen2.5-72B on AMD MI300X (BS=128), and (c) Mistral-7B on NVIDIA L40 (BS=128). The vertical lines mark the P5 and P95 values, bounding the central 90$\%$ of the distribution.}
    \label{fig:inter-node-var}
\end{figure*}

\begin{figure*}[htp]
    \centering
    \includegraphics[width=\textwidth]{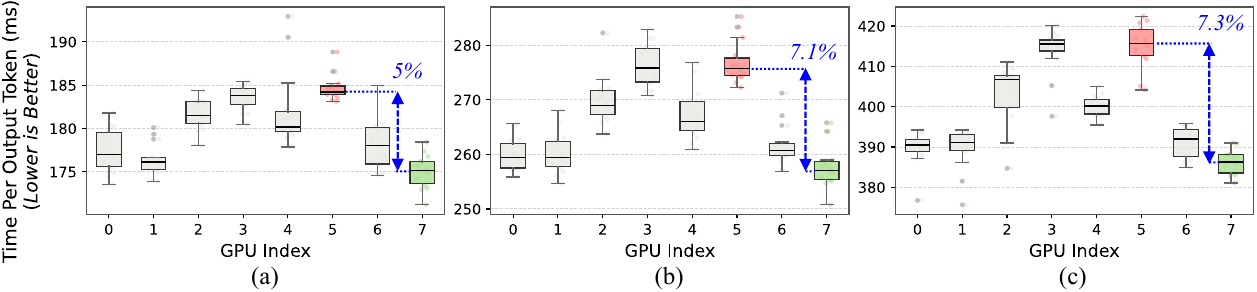}
    \caption{Intra-node TPOT variability across a NVIDIA 8xL40 node for DeepSeek-R1-Qwen-7B at batch sizes (a) 32, (b) 64, (c) 128. The median deviation between the best (GPU 7) and the worst (GPU 5) increases with batch size, showing that larger batches widen performance variability across GPUs within the same node.}
    \label{fig:intra-node-var}
\end{figure*}
\newpage
\section{Variability in GPU Systems}

\label{app:variability}

Modern GPU-accelerated systems exhibit significant performance \textit{variability}, so devices running identical workloads can produce measurably different execution times. This variability stems from several hardware and system level factors. At the hardware level, process variation during manufacturing produces differences in achievable clock frequencies and leakage characteristics across dies~\cite{sinha2022not, yoshida2022analyzing, tiwari2007recycle}. During runtime, dynamic voltage and frequency scaling (DVFS) continuously adjusts GPU clock frequency to stay within power and thermal limits. However, these adjustments depend on the thermal state of each GPU, which varies with physical placement and cooling mechanisms, causing GPUs in the same node to operate at different frequencies for the same workload. At the system level, OS scheduling, PCIe contention and virtualization layers further contribute to performance differences across GPUs~\cite{kurzynski2025lit, jeon2019analysis, chen2016baymax, chen2017prophet, xu2019characterization}.\\

\noindent To quantify the effects of variability on LLM workloads, we conducted a long-term measurement study across three GPU platforms: Amazon Tranium, AMD MI300X, and NVIDIA L40. We run state-of-the-art LLMs using vLLM with prompts from the ShareGPT~\cite{sharegpt} dataset. We report the variation in time per output token (TPOT), as decode iterations account for the majority of end-to-end latency in generative workloads~\cite{splitwise}. Figure~\ref{fig:inter-node-var} shows the aggregated TPOT across several nodes of a given platform. We observe that Amazon Tranium shows the tightest spread with samples within a range of only $1.44\%$, while NVIDIA L40 has the highest spread and shows 15.86$\%$ variability, with AMD MI300X falling in between. Moreover, even for GPUs within a single node show increasing variability with batch size. For example, Figure~\ref{fig:intra-node-var} shows that the difference between TPOT for GPU device 5 (slowest) and GPU device 7 (fastest) increases with batch size. This is because larger batches increase per-GPU power consumption and push each GPU closer to its thermal limits, this widens the variation in operating frequency on each GPU device. Furthermore, we note that the identity of the slowest and fastest GPUs remain identical across batch sizes, confirming that variability observed is persistent. This observed variability has direct implications for MoE inference. In expert-parallel deployments, experts are distributed across GPUs and each inference step requires all-to-all communication to route tokens to the selected expert. In these scenarios, the slowest GPU becomes the bottleneck or ``straggler" at every step and forces faster GPUs to idle at layer synchronization barriers. Thus, expert mapping strategies must account for variability in GPU performance to maximize MoE inference throughput.

%% file: sections/A1_Algorithm.tex
\newpage

\section{\papername's Expert Mapping Algorithm}
\label{app:algorithm}

This appendix describes the full expert mapping algorithm that is summarized in Section~\ref{sec:algo}. The expert mapping algorithm consists of two components: (i)~an algorithm that constructs an initial expert mapping, and (ii)~an iterative refinement algorithm that improves the initial mapping by evaluating alternate mappings. These two routines are combined in an outer loop that performs $K$ restarts and returns the lowest-scoring mapping across all restarts.

\vspace{0.5em}
\noindent\textbf{Notation.} We use the following notation:
\begin{itemize}
    \item $M_i^j$: the expert-to-GPU mapping at the beginning of the $j^{\text{th}}$ iteration of the $i^{\text{th}}$ restart. $M_i^0$ is the initial mapping for restart $i$, and $M_i^m$ is the final mapping.
    \item $T$: the expert utilization trace collected in Step-1. $T[t][e]$ is the number of tokens routed to $e$ at time step $t$.
    \item $C_g(\cdot)$: the per-GPU latency cost function from the variability profile in Step-2. 
    \item $u$: per-expert mean utilization derived from $T$.
    \item $S(M)$: the score defined in Section~\ref{sec:algo}, Equation~\ref{eq:score} that is used to evaluate mappings.
    \item $\textsc{Swap}(M, e_a, e_b)$: the mapping obtained after the positions of experts $e_a$ and $e_b$ are exchanged in $M$.
    \item $M \cup \{e \!\to\! g\}$: the mapping obtained by adding the assignment of expert $e$ to GPU $g$ in $M$.
\end{itemize}

\subsection{Initial Mapping Construction}

Algorithm~\ref{alg:init} builds an initial mapping $M_i^0$ greedily for restart $i$. Experts are sorted from most to least mean utilization and inserted one at a time onto whichever GPU yields the lowest score on the partial mapping. This ordering reflects the intuition that the heaviest experts have the largest impact on latency and should be placed first. To ensure that subsequent restarts refine a different initial mapping, the utilizations are perturbed by 20\% noise on each restart except the first.

\begin{algorithm}[ht]
\caption{\textsc{Initialmapping}\,$(u, i, T, C)$}
\label{alg:init}
\begin{algorithmic}[1]
\Require utilizations $u$, restart index $i$, trace $T$, curves $C$
\Ensure initial mapping $M_i^0$
\State $M_i^0 \gets \emptyset$ \textcolor{blue}{\textit{ // start with all GPUs empty}}
\If{$i > 1$} \textcolor{blue}{\textit{ // diversify across restarts}}
    \State Add 20\% noise to each utilization in $u$
\EndIf
\State Sort experts by $u$, highest first
\For{each expert $e$ in sorted order}
    \State $g^\star \gets \arg\min_{g}\; S(M_i^0 \cup \{e \!\to\! g\})$
    \State $M_i^0 \gets M_i^0 \cup \{e \!\to\! g^\star\}$ \textcolor{blue}{\textit{ // heaviest first}}
\EndFor
\State \Return $M_i^0$
\end{algorithmic}
\end{algorithm}

\subsection{Iterative Refinement}

Algorithm~\ref{alg:refine} improves an initial mapping $M_i^0$ by swapping the pair of experts which creates the largest reduction in $S$ at each iteration. The loop terminates when no swap reduces the score by more than $0.1\%$, indicating the mapping is stable.
\begin{algorithm}[ht]
\caption{\textsc{Refine}\,$(M_i^0, T, C)$}
\label{alg:refine}
\begin{algorithmic}[1]
\Require initial mapping $M_i^0$, trace $T$, curves $C$
\Ensure refined mapping $M_i^m$
\State $M \gets M_i^0$
\Repeat
    \State $s_{prev} \gets S(M)$; \quad $best \gets \text{None};\quad bestDrop \gets 0$
    \For{each pair $(e_a, e_b)$ on different GPUs in $M$}
        \State \textcolor{blue}{\textit{// try all cross-GPU swaps}}
        \State $M' \gets \textsc{Swap}(M, e_a, e_b)$
        \State $drop \gets s_{prev} - S(M')$
        \If{$drop > bestDrop$}
            \State $bestDrop \gets drop$
            \State $best \gets (e_a, e_b)$
        \EndIf
    \EndFor
    \If{$best = \text{None}$} \textbf{break}
        \State \textcolor{blue}{\textit{// no swap improves score}}
    \EndIf
    \State $M \gets \textsc{Swap}(M, best)$ \textcolor{blue}{\textit{ // commit best swap}}
\Until{$bestDrop\,/\,s_{prev} < 0.001$} \textcolor{blue}{\textit{ // converged}}
\State $M_i^m \gets M$
\State \Return $M_i^m$
\end{algorithmic}
\end{algorithm}

\subsection{Outer Loop: $K$ Restarts}

Because the greedy initialization and pairwise refinement are both heuristic, the final mapping depends on the starting point. To mitigate this, Algorithm~\ref{alg:outer} runs the full pipeline from $K$ different initial mappings and returns the lowest-scoring result among them. The first restart uses the unperturbed utilizations; subsequent restarts add 20\% noise to diversify their starting points (via Algorithm~\ref{alg:init}).

\begin{algorithm}[htp]
\caption{\textsc{\papername-Place}\,$(K, u, T, C)$}
\label{alg:outer}
\begin{algorithmic}[1]
\Require restart count $K$, utilizations $u$, trace $T$, curves $C$
\Ensure best expert-to-GPU mapping $M^\star$
\State $M^\star \gets \text{None};\quad s^\star \gets \infty$
\For{$i = 1$ to $K$} \textcolor{blue}{\textit{ // independent restart}}
    \State $M_i^0 \gets \textsc{Initialmapping}(u, i, T, C)$
    \State $M_i^m \gets \textsc{Refine}(M_i^0, T, C)$
    \State $s \gets S(M_i^m)$
    \If{$s < s^\star$} \textcolor{blue}{\textit{ // keep best so far}}
        \State $s^\star \gets s;\quad M^\star \gets M_i^m$
    \EndIf
\EndFor
\State \Return $M^\star$
\end{algorithmic}
\end{algorithm}

\noindent \papername deploys the mapping $M^\star$ returned by Algorithm~\ref{alg:outer} 
in Step-4 (Section~\ref{subsec:design-overview}). All four algorithms run on the CPU.

%% file: sections/A2_AdditionalResults.tex
\newpage

\begin{figure*}[htp]
  \centering
  \includegraphics[width=\linewidth]{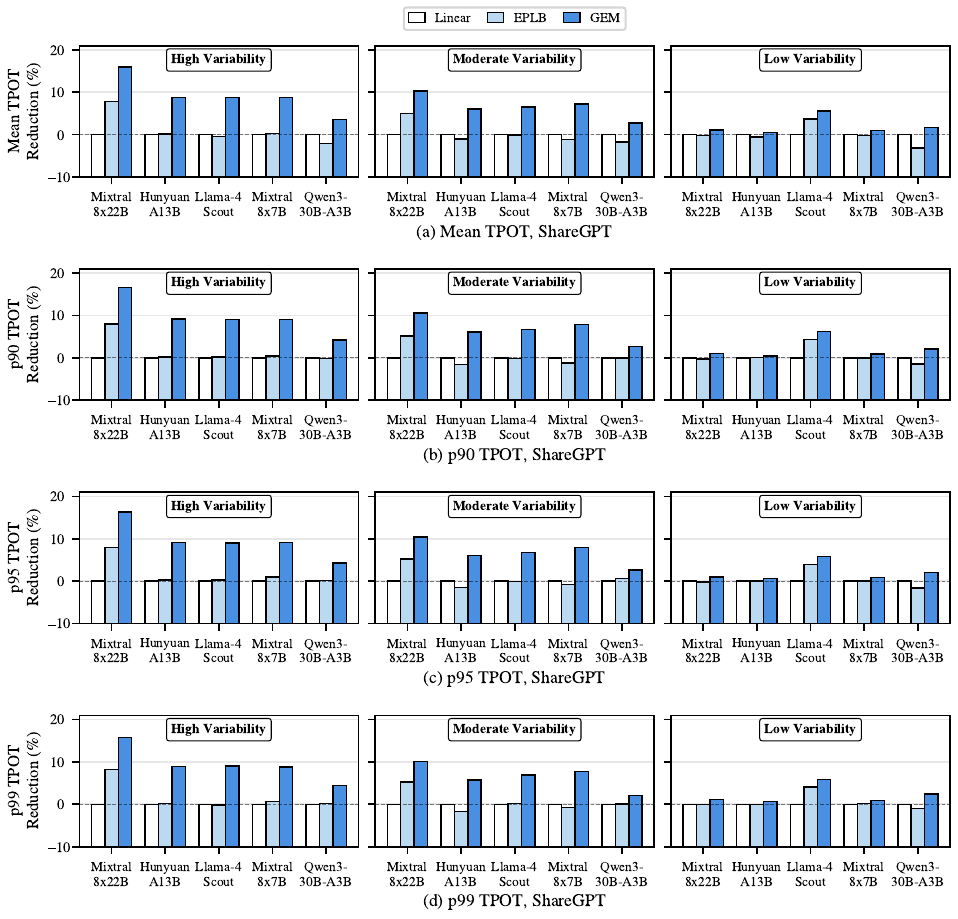}
  \caption{TPOT reduction relative to linear mapping on ShareGPT across all   five \moe models (higher is better). Rows report the TPOT statistic ((a)
  mean, (b) p90, (c) p95, (d) p99); columns report the variability setup.}
  \label{fig:tpot-percentiles-sharegpt}
\end{figure*}

\section{Tail-Latency Characterization}
\label{appendix:tail-latency}

MoE models are frequently deployed in streaming setups, where responses are sent to users one token at a time. When a few decode steps take longer to execute than most, the user notices visible stalls in token generation. These stalls hurt the user experience of the workload. To understand this, we report \papername's effect on tail latency, reporting the 90th, 95th, and 99th percentiles of time-per-output-token (TPOT) across both datasets and all three GPU variability setups.

\subsection{Metrics}
\label{appendix:tail-metrics}
Time-per-output-token (TPOT), is the elapsed time between two consecutive output tokens. We evaluate the TPOT distribution across 90th (p90), 95th (p95), and 99th (p99) percentiles as defined in Equation~(\ref{eq:pk-tpot}):
\begin{equation}
\label{eq:pk-tpot}
\text{p}k\ \text{TPOT} \;=\; Q_{k/100}\!\left(\Delta t\right),
\end{equation}
where $\Delta t$ denotes the inter-token latency across all tokens in the evaluation set, and $Q_{q}(\cdot)$ is the empirical $q$-quantile.

\begin{figure*}[!htp]
  \centering
  \includegraphics[width=\linewidth]{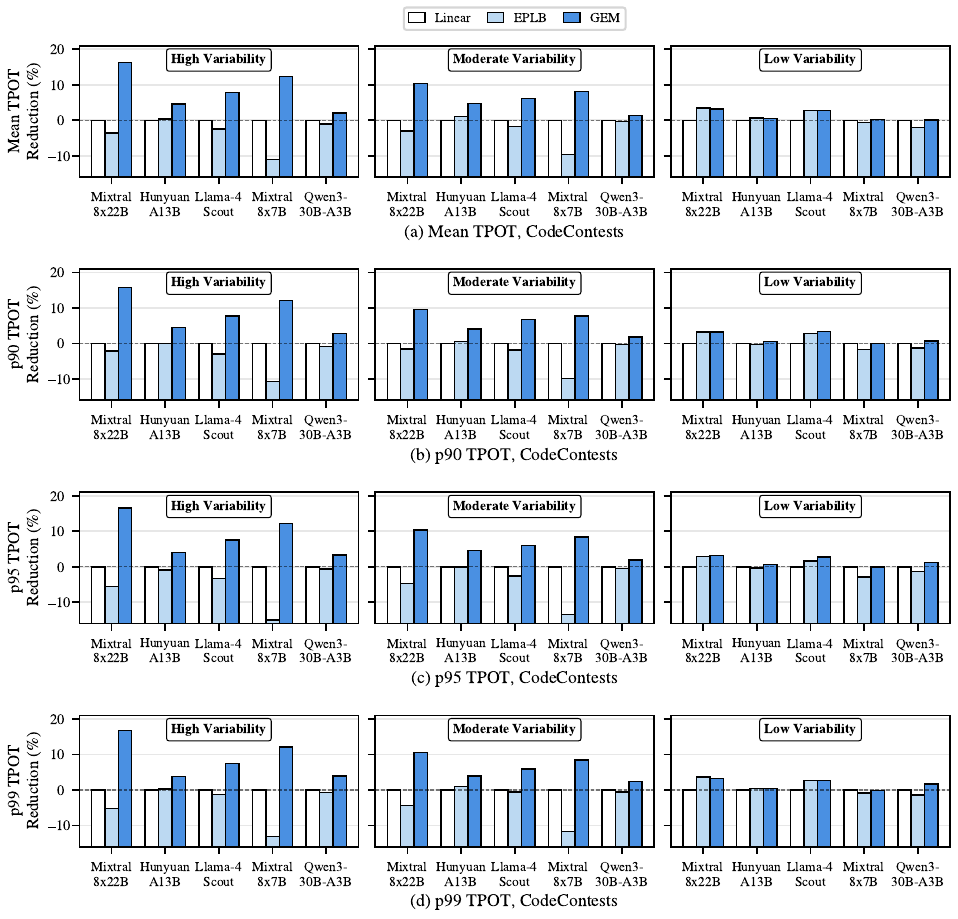}
  \caption{TPOT reduction relative to linear mapping on CodeContests across
  all five \moe models (higher is better). Rows report the TPOT statistic   ((a) mean, (b) p90, (c) p95, (d) p99); columns report the variability
  setup.}
  \label{fig:tpot-percentiles-codecontests}
\end{figure*}

\subsection{Tail-Latency Results}
\label{appendix:tail-results}

Figures~\ref{fig:tpot-percentiles-sharegpt}
and~\ref{fig:tpot-percentiles-codecontests} report TPOT reduction over linear mapping for ShareGPT and CodeContests. In each figure, the rows correspond to a TPOT statistic (mean, p90, p95, p99), and the three columns correspond to the high-variability, moderate-variability, and low-variability setups. We make two observations. \textit{First,} similar to end-to-end latency, \papername's benefit grows with the underlying hardware variability. On the high-variability setup, \papername reduces mean TPOT by {8.9\%} on average and by up to {16.8\%} in the best case. On the moderate-variability setup, it reduces mean TPOT by roughly {6\%} on average. On the low-variability setup,
it reduces mean TPOT by {1\%} to {2\%}. \textit{Second}, TPOT reductions are consistent across the distribution.  At a given variability level, the mean, p90, p95, and p99 reductions align to within half a percentage point. This indicates that \papername shifts the \textit{entire} per-token latency distribution, and improves both the typical streaming experience and the worst-case stalls.
